\documentclass[useAMS,usenatbib]{mn2e} \usepackage{graphicx}
\usepackage{epsfig} \usepackage{lscape} \usepackage{amsmath}
\usepackage{amssymb} \usepackage{aas_macros} \usepackage{bigints}
\usepackage{mathrsfs}
\usepackage{color}

\def \aj {AJ} 
\def \mnras {MNRAS} 
\def \apj {ApJ} 
\def \apjs {ApJS}
\def \apjl {ApJL} 
\def \aap {A\&A}

\def \prd {PhRvD}
\def \jcap{J. Cosm. Astro-Particle Phys.}

\def \Mpch {\, \mathrm{Mpc}\,h^{-1} \,}

\def\ls{\lower 2pt
  \hbox{$\;\scriptscriptstyle \buildrel<\over\sim\;$}} 
\def\gs{\lower
  2pt \hbox{$\;\scriptscriptstyle \buildrel>\over\sim\;$}}

\newcommand{\catrich}{R_{L^*}}

\newcommand{\mes}[3]{#1\pm_{#2}^{#3}}

\begin{document}

\title[BAO in the clustering of galaxy clusters] {An improved
  measurement of baryon acoustic oscillations from the correlation
  function of galaxy clusters at $z \sim 0.3$}

\author[Veropalumbo, Marulli, Moscardini, Moresco \& Cimatti]
       {A. Veropalumbo$^1$\thanks{E-mail:
           alfonso.veropalumbo@unibo.it}, F. Marulli$^{1,2,3}$,
         L. Moscardini$^{1,2,3}$, M. Moresco$^{1}$ and A. Cimatti$^1$
         \\ $^1$Dipartimento di Fisica e Astronomia, Universit\`a di
         Bologna, viale Berti Pichat 6/2, I-40127 Bologna,
         Italy\\ $^2$INAF - Osservatorio Astronomico di Bologna, via
         Ranzani 1, I-40127 Bologna, Italy\\ $^3$INFN - Sezione di
         Bologna, viale Berti Pichat 6/2, I-40127 Bologna, Italy}
       \pagerange{\pageref{firstpage}--\pageref{lastpage}}
       \pubyear{2013} \maketitle

\label{firstpage}

\begin{abstract}

We detect the peak of baryon acoustic oscillations (BAO) in the
two-point correlation function of a spectroscopic sample of $25226$
clusters selected from the Sloan Digital Sky Survey.  Galaxy clusters,
as tracers of massive dark matter haloes, are highly biased
structures.  The linear bias $b$ of the sample considered in this
work, that we estimate from the projected correlation function, is
$b\,\sigma_8=1.72\pm0.03$.  Thanks to the high signal in the
cluster correlation function and to the accurate spectroscopic
redshift measurements, we can clearly detect the BAO peak and
determine its position, $s_p$, with high accuracy, despite the
relative paucity of the sample. Our measurement,
$s_p=104\pm7\Mpch$, is in good agreement with previous estimates
from large galaxy surveys, and has a similar uncertainty. The BAO
measurement presented in this work thus provides a new strong
confirmation of the concordance cosmological model and demonstrates
the power and promise of galaxy clusters as key probes for
cosmological applications based on large scale structures.
\end{abstract}

\begin{keywords}
  cosmology: observations -- galaxy clustering -- large-scale
  structure of the Universe
\end{keywords}


\section{Introduction}
\label{s|intro}

The clustering of cosmic structures is one of the most powerful tools
to constrain cosmology. In particular, the signal of the baryon
acoustic oscillations (BAO) in the two-point correlation function acts
as a standard ruler, providing geometric cosmological constraints.
The accuracy in the determination of the position of the BAO peak
depends mainly on statistical uncertainties. By now the most accurate
measurements have been obtained with large spectroscopic samples of
galaxies \citep[e.g.][]{eisenstein2005, cole2005, percival2007,
  percival2010, sanchez2009, kazin2010, beutler2011, blake2011,
  pad2012, anderson2012, anderson2014} at low redshifts,
  $z<1$, and with Ly$\alpha$ forest in quasar spectra at higher
  redshifts \citep[e.g.][]{slosar2013, delubac2014}.

In recent analyses also galaxy clusters have been considered as probes
for the large scale matter distribution
\citep{angulo2005}. As tracers of the biggest collapsed
structures, they are more strongly clustered than galaxies.
Measurements of the two-point correlation function of galaxy clusters
have provided the first weak detections of the BAO peak.  
  \citet{estrada2009} and \cite{hutsi2010} measured, respectively, the
  two-point correlation function and the power spectrum of the MaxBCG
  photometric catalogue, consisting of $\sim 14000$ galaxy clusters
  \citep{koester2007}. Both works claimed a BAO detection with a
  significance of $1.5 < \sigma < 2$. Using a similar number of
  objects and in an more extended redshift range, \citet{hong2012}
  detected the BAO peak in the two-point correlation function of the
  spectroscopic cluster catalogue provided by \citet{wen2009}, with a
  confidence of $1.8\,\sigma$.

In this paper we present new measurements of the clustering of galaxy
clusters, up to the BAO scale, using the largest spectroscopic sample
currently available. As we will show, the BAO peak is clearly detected
at a scale $s_p \approx 105\,\mathrm{Mpc}\,h^{-1}$.

The paper is organized as follows. In \S \ref{s|data} we describe the
selected cluster sample used for this work, while the data analysis is
outlined in \S \ref{s|analysis}. In \S \ref{s|results} we present our
clustering measurements, we derive cosmological constraints from
  the position of the BAO peak and we compare them to previous
  studies. In \S \ref{s|disc} we compare the cluster clustering with
  the clustering of Luminous Red Galaxies (LRG), and we investigate the
  impact of photometric redshift errors.  Finally, in \S
\ref{s|conclu} we summarize our results.


\begin{figure*}
\centering
  \includegraphics[scale=0.9]{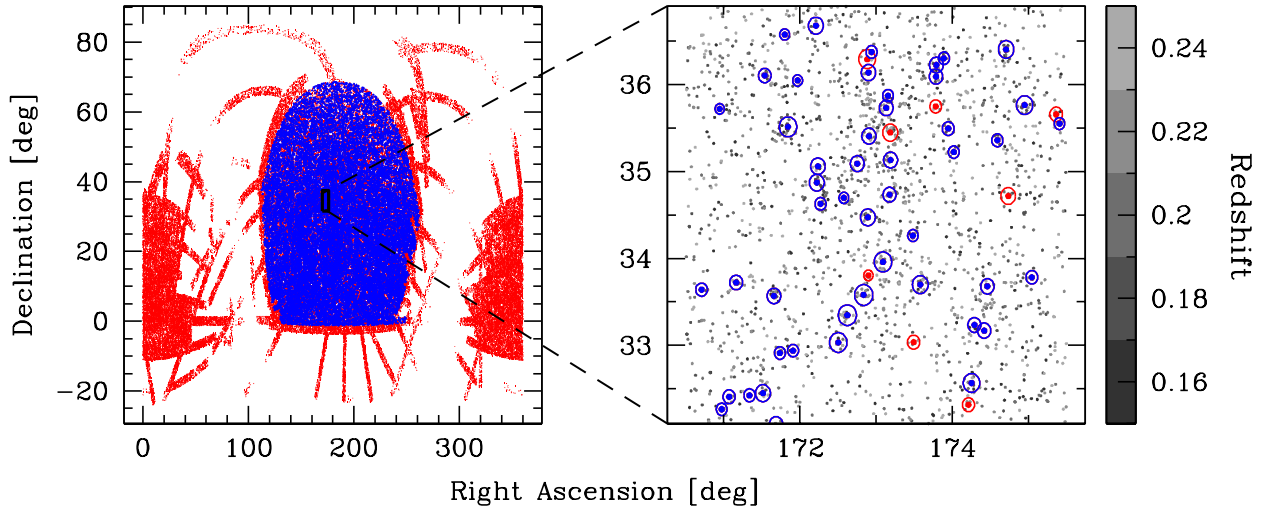} 
   \caption{{\em Left panel}: the angular distribution of
       galaxy clusters from the spectroscopic (blue dots) and
       photometric (red dots) samples analysed in this work.  {\em
         Right panel}: zoomed region of $5\times5$ square
       degrees. Grey points show the angular positions of galaxies
       from the SDSS DR8 photometric sample, selected in the redshift
       shell $0.15 < z < 0.25$, as indicated by the colour map. Blue
       and red circles represent the angular projection of the cluster
       radii $r_{200}$, from the spectroscopic (blue) and photometric
       (red) samples, estimated using our fiducial cosmology.}
  \label{f|data}
\end{figure*}


\section{Data}
\label{s|data}

We consider the spectroscopic cluster sample provided by
\citet*{wen2012} (WHL12), that has been extracted from the Sloan
Digital Sky Survey (SDSS) III \citep{aihara2011}. The cluster
candidates are identified by deprojecting the transversal
overdensities, using the information on photometric
redshifts. Clusters are included in the sample if they satisfy two
conditions: i) $N_{200}\geq 8$, where $N_{200}$ is the number of
galaxy members inside the radius $r_{200}$, at which the average
density is $200$ times the background density, and ii) $R_{L^*}\geq
12$, where $R_{L^*}$ is the ratio between $L_{200}$, the
\textit{r}-band luminosity inside $r_{200}$, and $L^*$, the
characteristic \textit{r}-band luminosity of galaxies
\citep[see][]{blanton2003}.  The cluster centre is determined by the
position of the brightest cluster galaxy (BCG), while its photometric
redshift is the median value of the photometric redshifts of its
galaxy members. A spectroscopic redshift is then assigned to a cluster
if it has been measured for its BCG. 

The total number of detected clusters in the whole photometric sample
is $132683$, in the redshift range $0.05<z<0.8$.  The detection rate
increases with the cluster mass. Using X-ray and weak-lensing
measurements available for a subsample of clusters, WHL12 showed that
the sample is complete for $M_{200}\gtrsim2\cdot10^{14} M_{\odot}$ in
the redshift range $0.1<z<0.42$, while the detection rate decreases
down to $\sim75 \%$ for the minimum mass of the sample, $M_{200} =
6\cdot10^{13} M_{\odot}$ (see WHL12 for more details on the detection
algorithm adopted).  For this work, we use a subsample of clusters
extracted from the WHL12 spectroscopic sample. Specifically, we
consider the complete spectroscopic cluster sample from the Northern
Galactic Cap, with measured redshifts in the range
$0.1<z<0.42$. Moreover, we use only the SDSS stripes with at least
$50\%$ of the clusters with spectroscopic redshift assigned. This is
to obtain the largest contiguous area and to minimize possible
selection effects. The final number of objects in our selected sample
is $25226$.  The left panel of Fig.~\ref{f|data} shows the
  angular distribution of the spectroscopic cluster sample (blue dots)
  analysed in this work, compared to the entire photometric sample
  (red dots), while the right panel shows a zoomed $5\times5$ square
  degrees region, where grey dots represent galaxies from the SDSS DR8
  photometric survey, and blue and red circles represent the angular
  projection of the cluster radii $r_{200}$, from the spectroscopic
  (blue) and photometric (red) samples, estimated using our fiducial
  cosmology.

The main properties of the spectroscopic sample used for this
  work are summarized in Table \ref{t|data}. The photometric sample
is used to compute the sampling rate, as described in \S
\ref{s|weights}. In Fig.~\ref{f|zdist} we show the redshift
  distribution of the selected spectroscopic clusters. The bimodal
  shape is due to the presence of two main spectroscopic targets in
  SDSS-II: the main sample that peaks around $z\sim0.12$
  \citep{strauss2002}, and the LRG sample that covers the redshift
  range $0.2 < z < 0.5$ \citep{eisenstein2001}.


\section{Analysis}
\label{s|analysis}

\subsection{Two-point correlation function in real-space and redshift-space} 

We estimate the redshift-space two-point correlation function,
$\xi(s)$, using the \citet{landy1993} estimator:
\begin{equation}
\xi(s) \, = \, \frac{1}{RR(s)}\times\left[
  DD(s)\frac{n_r^2}{n_d^2}-2DR(s)\frac{n_r}{n_d} +RR(s) \right] \, ,
\label{e|lsext}
\end{equation}
where $DD(s)$, $DR(s)$ and $RR(s)$ are the numbers of weighted
data-data, data-random and random-random pairs within a separation
$s\pm\Delta s /2$, where $\Delta s$ is the bin size, and $n_r$ and
$n_d$ are the weighted number density of random and cluster sample,
respectively. To compute comoving distances and bias (see \S
\ref{ss|df}), we assume a flat $\Lambda$ cold dark matter
(CDM) model with the mass density parameter $\Omega_M=0.3$,
the baryon density parameter $\Omega_b=0.045$, the Hubble constant
$H_0=70 \, \mathrm{km} \, \mathrm{s}^{-1} \, \mathrm{Mpc}^{-1}$, the
primordial perturbation spectral index $n_s=1$, and the linear power
spectrum amplitude $\sigma_8=0.8$.

To derive the real-space clustering, we measure the projected
correlation function:
\begin{equation}
w_p(r_p) \, = \, \int_0^{\pi_{max}} \mathrm{d} \pi^{\prime}
\xi(r_p,\pi^{\prime}) \, ,
\label{e|wp}
\end{equation}
where $\xi(r_p,\pi)$ is the measured two-point correlation function in
the directions perpendicular, $r_p$, and parallel, $\pi$, to the
line-of-sight.  The real-space two-point correlation function,
$\xi(r)$, is then obtained from $w_p$ assuming a power-law model,
$\xi(r)=(r/R_0)^{-\gamma}$, where $R_0$ and $\gamma$ are the
correlation length and the power-law index, respectively.  With the
above assumption, the relation between $\xi$ and $w_p$ can be derived
analytically:
\begin{equation}
w_p(r_p) \, = \, r_p \left( \frac{R_0}{r_p} \right)^{\gamma}
\frac{\Gamma(\frac{1}{2})\Gamma(\frac{\gamma-1}{2})}{\Gamma(\frac{\gamma}{2})}
\, ,
\label{e|wpan}
\end{equation}
where $\Gamma$ is the Euler's gamma function.


\begin{figure}
\includegraphics[width=0.45\textwidth]{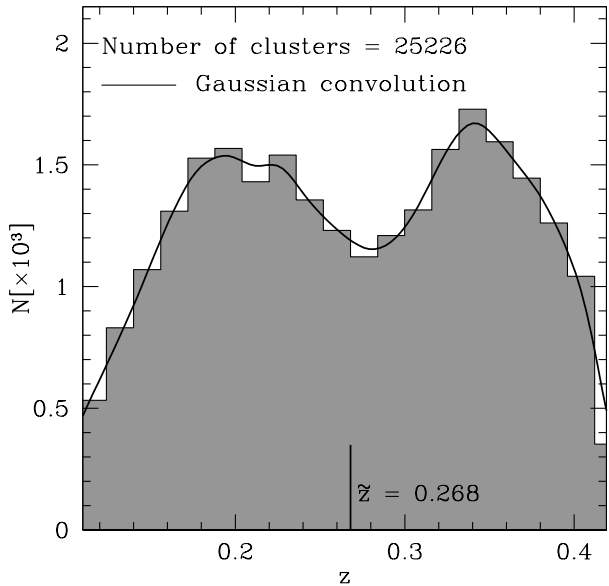}
  \caption{The redshift distribution of the selected galaxy clusters
    (histogram). The solid line shows the smoothed redshift
    distribution obtained adopting a Gaussian filter. See WHL12 for
    further details on the redshift distribution of the galaxy samples
    used to detect the clusters. }
  \label{f|zdist}
\end{figure}

\begin{table}
\centering
\caption{Main properties of the spectroscopic cluster sample
  used for this work. See \S \ref{s|data} for more details.}
\label{t|data}
\renewcommand{\arraystretch}{1.3}
\begin{tabular}{c|c}
\hline
\hline
$\#$ objects & $25226$ \\
Area $[\mathrm{deg}^2]$ & $\sim8400$ \\
$z$ range & $0.1 < z < 0.42$ \\
$\tilde{z}$ & $0.268$ \\
$R_{L^*}^{min}$ & $12$ \\ 
$M_{min}$ & $6\cdot10^{13} M_{\odot}$ \\
\hline
\hline
\end{tabular}
\end{table}


\subsection{Random sample}
\label{s|rancat}

To measure the two-point correlation function of our sources, we have
to construct a sample of randomly distributed objects (see
Eq.~\ref{e|lsext}), taking into account the selection function of the
sample. As a fair approximation, we can factorise the random sample
distributions into the angular and redshift components separately.

The angular mask is reconstructed with the software {\small MANGLE}
\citep{swanson2008}. Using the SDSS coordinates system
$(\lambda,\eta)$, we decompose the angular distribution of clusters in
rectangular elements of equal area, that are then randomly filled. We
do not apply any weights to take into account sector completeness when
creating the random sample.

We assign redshifts to the random objects sampling the mean redshift
distribution of the catalogue. The latter has been obtained grouping
the data in $100$ redshift bins and smoothing the distribution with a
Gaussian kernel three times larger than the bin size. Reducing the
value of this parameter has the effect to lower the clustering signal
in the radial direction. The impact of this effect is however
negligible, considering the estimated uncertainties in our
measurements. To minimize the effect of shot noise, we construct a
random sample ten times denser than the cluster sample.
Fig.~\ref{f|zdist} shows the redshift distribution of the cluster
sample (histogram) and the smoothed distribution (solid line) used for
the construction of the random sample.


\subsection{Weights}
\label{s|weights}

In this analysis, we apply three different weights to correct for i)
the effects of a mass-dependent detection rate in the cluster
selection algorithm, ${W}_{M_{200}}$ (see WHL12), and for ii) the
spectroscopic sampling rate, as a function of the cluster richness,
${W}_{N_{200}}$, and of the stripe location, ${W_S}$, separately.  We
derive the above quantities directly from the data, comparing the
photometric and spectroscopic cluster samples. For each cluster, the
total weight assigned is:
\begin{equation}
w_i(\catrich,N_{200},stripe) = {W}_{M_{200}}^{-1} \cdot
{W}_{N_{200}}^{-1} \cdot {W}_S^{-1} \, .
\label{e|weight}
\end{equation}
The net effect is to increase the number of low mass structures, whose
sampling rate is lower with respect to the more massive structures, in
the spectroscopic sample. This slightly reduces the clustering
normalization, up to $\sim 10\%$. On the contrary, we find that the
BAO peak position is not affected by the details of the weighting
scheme adopted.


\subsection{Error estimates}

The errors on the clustering measurements are estimated with the
jackknife method \citep[see e.g.][]{norberg2009}.  The covariance
matrix for the jackknife estimator is:
\begin{equation}
C_{ij} \, = \, \frac{N_{sub}-1}{N_{sub}} \sum_{k=1}^N
(\xi_i^k-\bar{\xi}_i)(\xi_j^k-\bar{\xi}_j) \, ,
\label{e|cov}
\end{equation}
where $\xi_i^k$ is the value of the correlation function at the
\textit{i}-th bin for the \textit{k}-th subsample, and $\bar{\xi}_i$
is the mean value of the subsamples.

We construct $N_{sub}=140$ resamplings of our cluster catalogue
by dividing the original sample in $N_{sub}$ regions (i.e. $5$
subvolumes for each of the $28$ SDSS stripes considered) and excluding
recursively one of them. Increasing the number of subregions
  provides a less scattered estimate of the covariance matrix. As
  verified directly, the value of $N_{sub}$ adopted here is large
  enough to assure the convergence of the results.  We extensively
  test the jackknife algorithm exploited in this work using the
  {\small LasDamas} mock catalogues \citep{mcbride2009}, finding that
  the quoted errors are conservative estimates.


\subsection{Models}
\label{ss|df}

In the following sections, we describe the models used to
  derive clustering parameters and cosmological constraints from the
  projected correlation function and the BAO peak.  The analysis is
  performed applying a Monte Carlo Markov Chain (MCMC) technique,
  using the full covariance matrix. We adopt a standard likelihood,
  $\mathscr{L}\propto\exp(-\chi^2/2)$, where the function $\chi^2$ is
  defined as follows:
  \begin{equation}
    \chi^2 = \sum_{i=0}^{i=n} \sum_{j=0}^{j=n} (\xi_i - \hat{\xi}_i)
    C^{-1}_{ij} (\xi_j - \hat{\xi}_{j}) \, ,
    \label{e|chi2}
  \end{equation} 
  where $\xi_i$ is the correlation function measured in the
  \textit{i}-th bin, $\hat{\xi}_i$ is the model and $C^{-1}_{ij}$ is
  the inverted covariance matrix.


\subsubsection{The cluster bias}
\label{sss|cbm}

To measure the bias factor, $b$, we model the projected
  correlation function assuming a linear biasing model,
\begin{equation}
  w_p(r_p)=b^2 w_p^{DM}(r_p) \, ,
\label{e|blin}
\end{equation} 
where $w_p^{DM}$ is the DM projected correlation function
\citep[e.g.][]{marulli2013}. When assessing the bias through
Eq.~\ref{e|blin}, the upper limit of the integration in
Eq.~\ref{e|wp}, $\pi_{max}$, has to be fixed. The impact of this
parameter choice is not significant, considering the estimated
uncertainties. Nevertheless, a finite value of $\pi_{max}$ introduces
unavoidable systematic errors, as the effect of redshift-space
distortions (RSD) can not be entirely washed out by the
integration. The net effect is a spurious scale-dependence in the
estimated bias. To minimize the impact of such a systematics, instead
of using Eq.\ref{e|blin} we model directly the projected correlation
function as follows:
\begin{equation}
  w_p(r_p) \, = \, b^2 \int_0^{\pi_{max}} \mathrm{d}
  \pi^{\prime} \xi^{\rm DM}(r_p,\pi^{\prime}) \, ,
  \label{e|wpteo}
\end{equation}
where the value of $\pi_{max}$ is the same as the one used to measure
$w_p(r_p)$ and $\xi^{\rm DM}(r_p,\pi)$ is the redshift-space DM
correlation function in the directions perpendicular and parallel to
the line of sight. RSD are introduced with the {\em dispersion model}
\citep{kaiser1987, hamilton1992, davis1983}, following
\citet{marulli2012}. The linear DM correlation function is obtained by
Fourier transforming the matter power spectrum computed with the
software {\small CAMB} \citep{lewis2002}. The linear RSD parameter is
estimated assuming a $\Lambda{\mathrm{CDM}}$ cosmology, i.e.
$\beta=\Omega_M(z)^{\gamma}/b$, with $\gamma=0.545$.

As extensively tested, this method is able to compensate for
  the effect of RSD when integrating up to a finite value of
  $\pi_{max}$, providing an approximately scale-independent bias in
  the range of scales considered.


\subsubsection{Cosmological constraints from the BAO peak}
\label{sss|baomodel}

In this section, we descibe two different methods to detect
  the BAO peak and extract cosmological information.  Results obtained
  with both the methods are presented in \S \ref{s|bao}.


\subsubsection*{Empirical model}
\label{sss|analyticalmodel}

We consider an empirical model similar to the one proposed by
  \cite{sanchez2012}, which is used to interpolate the function $\xi(s)$
  at the BAO scales:
\begin{equation}
\xi(s) \,=\,
B+\left(\frac{s}{s_0}\right)^{-\gamma}+\frac{N}{\sqrt{2\pi\sigma^2}}
\exp\left(-\frac{(s-s_m)^2}{2\sigma^2}\right)\, ,
\label{e|sanchezmodel}
\end{equation}
where the parameters $s_0$ and $\gamma$ model the shape of the
correlation at small scales, $B$ takes into account a possible
negative correlation at large scales, and $s_m$, $\sigma$, and $N$ are
the parameters of the Gaussian function used to model the BAO feature.
We note that the true BAO peak position, $s_p$, is shifted to smaller
scales with respect to the Gaussian median value $s_m$.

The empirical model given by Eq.~\ref{e|sanchezmodel} can be
  used to accurately detect the BAO peak position.  To directly
  compare our measurements with previous studies, we compute also the
  dimensionless variable:
  \begin{equation}
  y_s = \frac{r_s}{D_V} \, , \label{e|ys} 
  \end{equation}   
that results to be independent
  of the fiducial cosmology assumed to derive comoving coordinates
  \citep[see e.g.][]{sanchez2012}. The distance $D_V$ is defined as:
\begin{equation}
D_V = \left[(1+z)^2 D_A(z)^2\frac{cz}{H(z)}\right]^{\frac{1}{3}} \, ,
\label{e|DV}
\end{equation}  
where $D_A$ is the angular diameter distance and $H(z)$ is the Hubble
function.


\subsubsection*{Physical model}
\label{sss|teomodel}

To extract the full cosmological information embedded in the
  position of the BAO peak, we consider a theoretical model that
  includes the cluster bias, the effects of RSD and geometric
  distortions due to a possible incorrect assumption of the fiducial
  cosmology.  The adopted model is the following:
\begin{equation}
\xi_{cl}(s) \, = \, b^2 \, \left(1+\frac{2}{3}\beta+\frac{1}{5}\beta^2
\right) \xi_{DM} (\alpha s) \, ,
\label{e|bxidm}
\end{equation}
where $b$ is the linear bias factor, $\alpha$ is the ratio between the
test and fiducial values of $D_V$ and it is used to model geometric
distortions, and $\beta$ is the linear distortion parameter described
in \S \ref{sss|cbm}.  The non-linear DM correlation function,
$\xi_{DM}$, is computed using the software {\small MPTbreeze}
\citep{crocce2008}, based on the renormalized perturbation theory
\citep{crocce2006}.  This method has already been used
  in previous works aimed at extracting cosmological information from
  the position of the BAO peak \citep[see e.g.][]{eisenstein2005,
    beutler2011, blake2011}.

To compare with previous studies, we exploit this method to
  derive also other parameters such as $y_s$ (see also \S
  \ref{sss|analyticalmodel}), and the acoustic parameter $A(z)$,
  defined as follows:
\begin{equation}
A(z) \equiv \frac{100 D_V(z) \sqrt{\Omega_M h^2}}{cz} \, .
\label{e|Az}
\end{equation} 
This parameter results to be independent of $H_0$, since $D_V \propto
H_0^{-1}$ \citep[see e.g.][]{eisenstein2005,blake2011}.

\begin{figure}
  \begin{center}
    \includegraphics[width=0.9\linewidth]{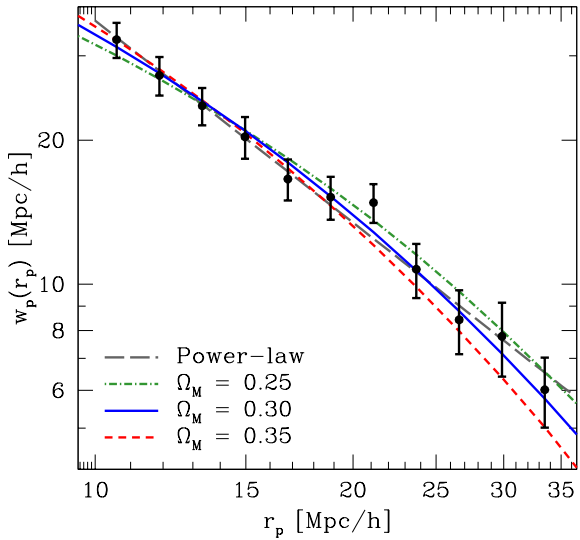}
    \caption{The projected correlation function of galaxy clusters
      (black dots). The dashed grey line shows the best-fit linear
      model defined by Eq.~\ref{e|wpan}, while the other three
        lines are the best-fit models obtained through
        Eq.~\ref{e|wpteo} for three different values of the mass
        density parameter, $\Omega_M=0.25,0.3,0.35$.}
    \label{f|bias}
  \end{center}
\end{figure}

\section{Results}
\label{s|results}

In this section, we present the main results of our
  analysis. We start focusing on the small scale clustering,
  estimating the linear bias from the projected correlation function
  at $r_p<30 \Mpch$. Then, we move to larger scales, detecting the BAO
  peak and extracting cosmological information. Finally, we compare
  our measurements with previous studies.


\begin{figure*}
  \begin{tabular}{p{0.45\textwidth} p{0.51\textwidth}}
    \vspace{0pt}
    \includegraphics[width=\linewidth]{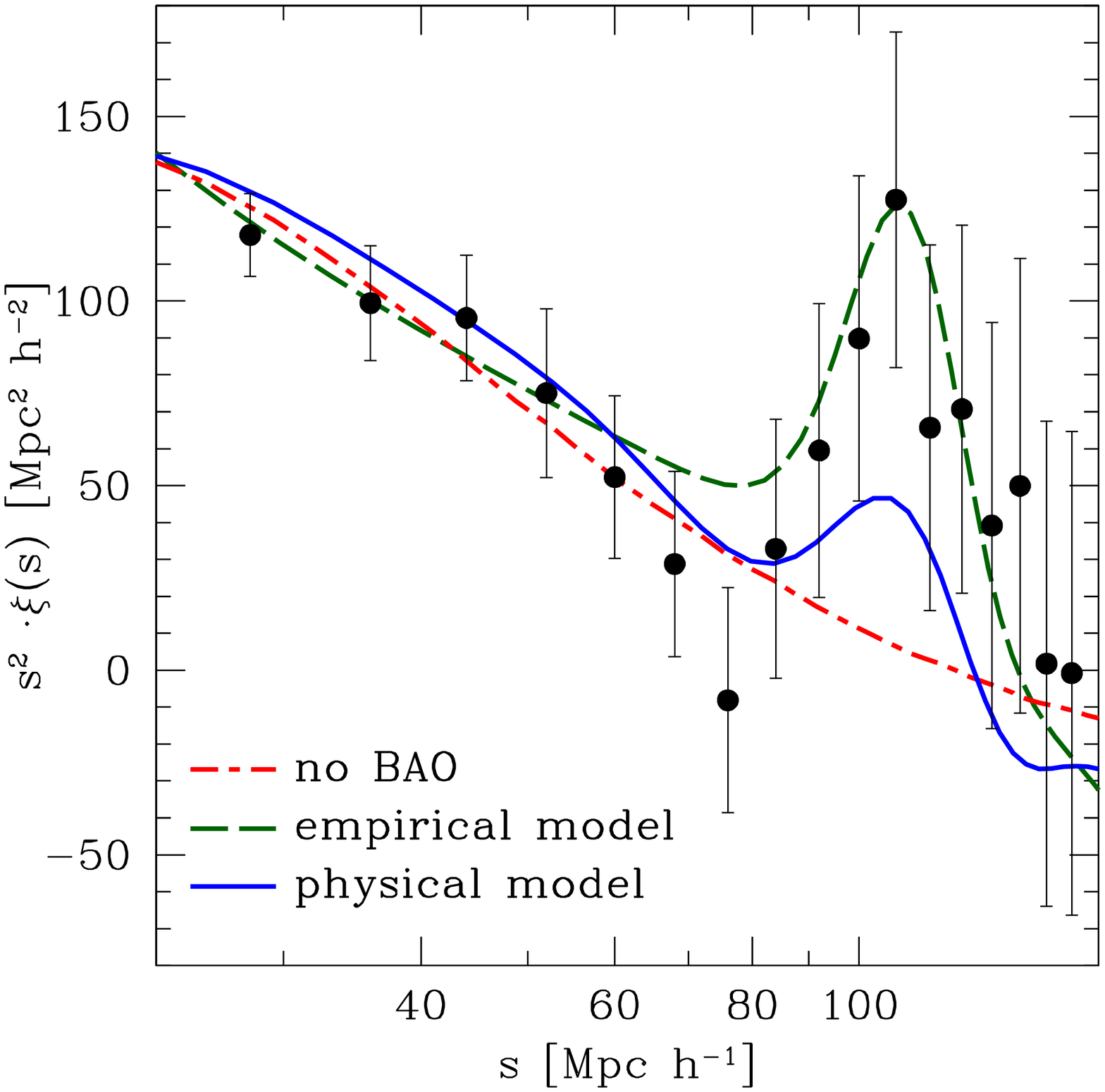} &
    \vspace{0pt}
    \includegraphics[width=0.45\textwidth]{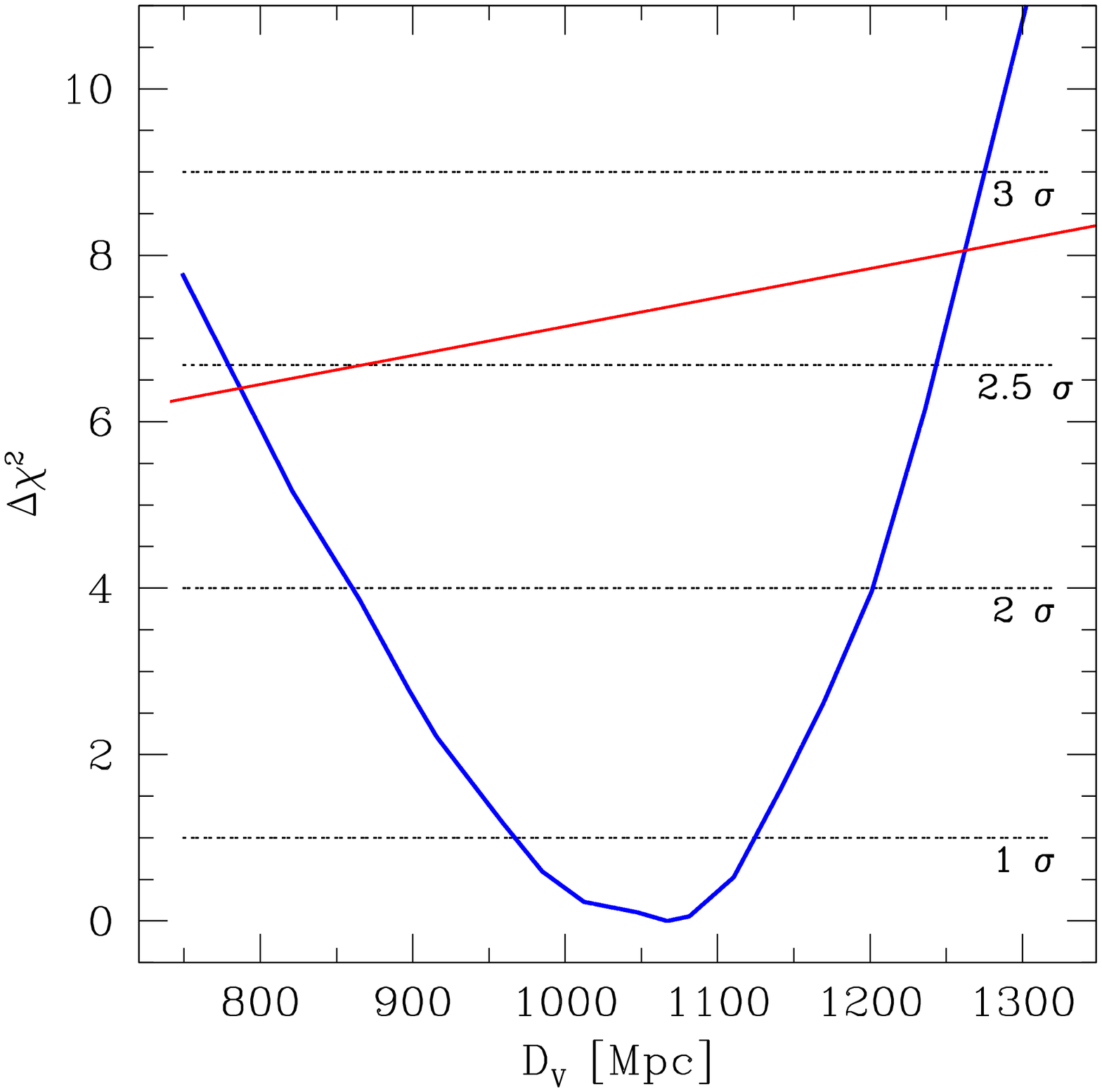}
  \end{tabular}
  \caption{{\em Left panel}: the redshift-space two-point correlation
    function of galaxy clusters (black dots), multiplied by $s^2$ to
    magnify the BAO peak; error bars are the square root
      of the diagonal elements of the covariance matrix, multiplied by
      $s^2$.  The dashed green line is the best-fit empirical
      model obtained through Eq.~\ref{e|sanchezmodel}. The blue line
      is the best-fit physical model given by Eq.~\ref{e|bxidm}, while
      the dot-dashed red line shows the no-BAO prediction, obtained
      with the fitting formula by \citet{eisenstein1999}.  {\em Right
      panel}: $\Delta \chi^2$ as a function of $D_V$, for the
      physical (blue line) and the no-BAO (red line) models. The BAO
      peak is detected with a $\sim 2.5 \sigma$ confidence level.}  
  \label{f|xi}
\end{figure*}

\subsection{Projected correlation function and bias}
\label{ss|pbias}

Fig.~\ref{f|bias} shows the projected correlation function,
$w_p(r_p)$, estimated through Eq.~\ref{e|wp}. The error bars are the
square root of the diagonal elements of the covariance matrix given by
Eq.~\ref{e|cov}, i.e. $\sigma_i \, = \, \sqrt{C_{ii}}$.  We derive the
correlation length, $R_0$, and the power-law index, $\gamma$, assuming
a power-law model for the real-space clustering, thus fitting the
projected correlation function using Eq.~\ref{e|wpan}.  The result of
the fit, obtained in the range of scales $10 < r_p
[\mathrm{Mpc}\,h^{-1}] < 30$, is shown by the dashed grey line.

In Eq.~\ref{e|wp} we set the upper limit of the integration to the
value $\pi_{max}=40 \, \mathrm{Mpc} \, h^{-1}$. We investigated the
impact of this assumption and of the scale range used for the fit,
repeating the procedure for different values of $\pi_{max}$ and of the
scale limits. We find that our results are only marginally affected by
these parameters. The maximum variation in $R_0$ and $\gamma$ is of
the order of $7\%$, when $\pi_{max}$ and the scale limits are changed
inside reasonable ranges (i.e. $20 < \pi_{max} [\Mpch] <
    60$, $ 5 < r_p [\Mpch] < 60$).

We estimate the linear bias parameter, $b$, using the method
  described in \S \ref{sss|cbm}. The DM correlation function is
  computed assuming the same fiducial cosmology used to measure
  comoving distances. The best-fit value of the bias, with 1$\sigma$
  uncertainties, is $b\,\sigma_8=1.72\pm0.03$, corresponding to a
  minimum $\chi^2$ value of $6.7$, with $10$ degrees of freedom. The
  best-fit values of $R_0$, $\gamma$ and $b\,\sigma_8$ are reported in
  Table~\ref{t|parameters}.

To investigate the impact of the mass density parameter, we
  repeat the same measurement for $\Omega_M=0.25$ and
  $\Omega_M=0.35$. The three best-fit models corresponding to the
  three assumed values of $\Omega_M$ are shown in Fig.~\ref{f|bias}
  with different lines, as indicated by the labels.  The
  measured $w_p(r_p)$ results to be only marginally affected by
  geometric distortions when changing $\Omega_M$, while this is not
  the case for the model. Therefore, the best-fit value of the bias
  does depend on $\Omega_M$ \citep[e.g.][]{marulli2012}. The best-fit
  values we obtain are the following: $b\,\sigma_8(\Omega_M=0.25)=1.55
  \pm 0.03$ and $b\,\sigma_8(\Omega_M=0.35)=1.88 \pm 0.04$. In
  particular, we find that the best-fit bias values derived for
  different $\Omega_M$ are the ones that keep the value of $\beta$
  approximately constant. The correspondent $\chi^2$ are:
  $\chi^2(\Omega_M=0.25)=8.6$ and $\chi^2(\Omega_M=0.35)=9.9$,
  respectively.  The minimum of $\chi^2$ is obtained for
  $\Omega_M=0.3$, thus favouring the fiducial cosmology assumed in
  this work. In the next section we will perform a more detailed
  analysis, modelling the large scale clustering and constraining
  directly $\Omega_M$, with a full MCMC method, finding consistent
  results.

We notice that our analysis shows a lower clustering with respect to
the estimates by \citet{estrada2009} and \citet{hong2012}. This is due
to the lower mass limit in our cluster sample, that results in a lower
bias.


\subsection{The BAO peak}
\label{s|bao}

\subsubsection{Fitting with the empirical model} 

The left panel of Fig.~\ref{f|xi} shows the redshift-space two-point
correlation function, $\xi(s)$, multiplied by $s^2$, in order to
magnify the BAO peak.  We start fitting the clustering data
  with the empirical model given by Eq.~\ref{e|sanchezmodel}, in the
  scale range $20 < s [\Mpch] < 180$, using a MCMC technique.  The
  result of the fit is shown by the dashed green line. The best-fit
  value of the peak position is $s_p=\mes{104}{6}{7} \Mpch$, after
  marginalizing over the other $5$ parameters of the model. When using
  linear instead of logarithmic binning, the BAO peak results slightly
  shifted to higher values. However, the effect is of the order of
  $2\%$, well below the estimated accuracy on the BAO peak position,
  that is of the order of $7\%$. We also fit the data with the same
  empirical model but without the Gaussian part. The $\Delta \chi^2$
  gives a confidence level for the full model between $2$ and
  $3\sigma$.

Our measurement is in good agreement with the previous detection by
\citet{hong2012}. Moreover, thanks to the higher cluster density in
our sample, that is larger by a factor of two, the uncertainty in the
position of the BAO peak is significantly lower.


\subsubsection{Fitting with the physical model}

We now fit the measured correlation function with the physical
  model described in \S \ref{sss|teomodel}.  Cosmological information
  is encoded in $\Omega_M h^2$, in the linear bias $b$, and in the
  shift parameter, $\alpha$, that traces geometrical distortions.  All
  the other cosmological parameters are kept fixed to the Planck
  values: $H_0=67.4 \, \mathrm{km \, s^{-1} \, Mpc^{-1}}$,
  $\Omega_b=0.02207h^2$, $n_s=0.96$ and $\sigma_8=0.83$
  \citep{planck2013}.

The best-fit parameters are summarized in
  Table~\ref{t|parameters}.  The reported values are the medians of
  the MCMC parameter distributions, while the $1 \sigma$ errors span
  from the $16th$ to the $86th$ percentiles.  The solid blue line in
  the left panel of Fig.~\ref{f|xi} shows the result of the fit
  obtained using the {\small MPTbreeze} software to estimate
  $\xi_{DM}(r)$, while the red one has been obtained using the fitting
  formula given by \citet{eisenstein1999} with no BAO.  As shown in
  the right panel of Fig.~\ref{f|xi}, the BAO feature is detected with
  a $\sim 2.5 \sigma$ confidence level, in agreement with what
  obtained with the empirical model. We achieve a distance measure of
  $D_V=\mes{1031}{92}{84}$.  Constraints on the distortion parameters
  $y_s$ are of the order of $7\%$, in good agreement with the value
  obtained with the empirical model in \S \ref{sss|analyticalmodel}.
  Fitting in the range $ 20 < s [\Mpch] < 180$, we obtain a $10\%$
  constraint on the mass density parameter, $\Omega_M h^2 =
  \mes{0.15}{0.02}{0.03}$, after marginalizing over the other two
  model parameters $\alpha$ and $b$ (see \S
  \ref{sss|teomodel}). Reducing the fitting range has the effect of
  slightly worsening the constraints.

\begin{figure}
  \center
  \includegraphics[width=0.45\textwidth]{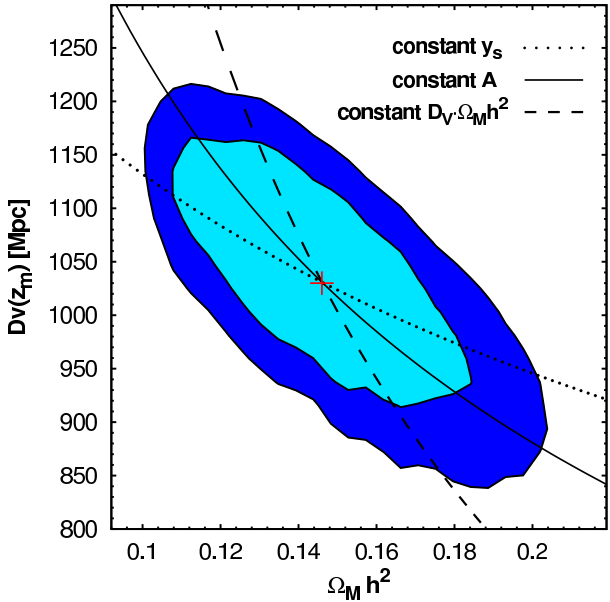}
  \caption{Marginalized probability contours at $1$ and $2
      \sigma$ for $\Omega_M h^2 - D_V$, obtained fitting
      Eq.~\ref{e|bxidm} in the scale range $20 < s [\Mpch] < 180$.
      The red cross marks the median values of the two parameters from
      the MCMC realizations.  The dotted line is obtained keeping
      fixed $y_s$ at the best-fit value, indicating a pure BAO-scale
      driven fit. The continuous line shows the same but for the
      parameter $A$ (Eq.~\ref{e|Az}). The dashed line is obtained
      keeping fixed the value of $D_V \Omega_M h^2$.}
  \label{f|cntDv}
\end{figure}

Fig.~\ref{f|cntDv} shows the $1$ and $2 \sigma$ marginalized
  probability contours in the $\Omega_M h^2 - D_V$ plane.  The dotted line
  indicates the points with constant $y_s$, i.e. it represents the
  degeneracy direction between parameters that would occur if the fit
  was driven by the BAO feature only. The dashed line shows the
  opposite case in which the fit is driven only by the shape of the
  two-point correlation function. As it can be seen, the orientation of
  the parameter degeneracy obtained in this work lies approximately in
  the middle between these two extremes, closely following the
  solid line of costant $A$ (Eq.~\ref{e|Az}). In Table
  \ref{t|parameters} we report the best-fit values of the cosmological
  parameters, as well as the estimated uncertainties derived from the
  MCMC analysis after marginalizing over all the free parameters of
  the fit. As it can be seen, the estimated value of $b$ is consistent
  with the one derived in \S \ref{ss|pbias} by fitting the projected
  correlation function at smaller scales.

\begin{table}
\centering
\caption{Best-fit parameters and 1 $\sigma$ uncertainties
    obtained from the projected and redshift-space two-point
    correlation functions of the selected spectroscopic cluster
    sample. For more details see \S \ref{sss|cbm} and \S
    \ref{sss|baomodel}.}
\label{t|parameters}
\renewcommand{\arraystretch}{1.3}
\begin{tabular}{c|l|c}
\hline
\hline
Statistics & Parameters & Best-fit values, $1\sigma$ uncertainties 
\\
\hline
& $R_0$ $[\Mpch]$ & $11.4 \pm 0.4$ \\
$w_p(r_p)$ & $\gamma$ & $2.3 \pm 0.1$ \\
& $b\,\sigma_8$ &$1.72 \pm 0.03$ \\
\hline
& $s_p$ $[\Mpch]$ & $\mes{104}{6}{7}$ \\
& $D_V(\tilde{z})$ $[\mathrm{Mpc}]$ & $\mes{1031}{92}{84}$ \\
$\xi(s)$ & $y_s$ & $\mes{0.147}{0.008}{0.010}$ \\
& $A(\tilde{z})$ & $\mes{0.48}{0.03}{0.03}$ \\
& $\Omega_M\,h^2$ & $\mes{0.15}{0.02}{0.03}$ \\
& $b\,\sigma_8$ &$\mes{1.6}{0.2}{0.1}$ \\ 
\hline
 \hline
\end{tabular}

\end{table}


\subsection{Comparison with previous BAO measurements}

To compare our measurements with similar results in the literature, we
estimate the dimensionless variable $y_s=s_p/D_V$ as described in \S
\ref{ss|df}, assuming the median redshift of the sample, $\tilde{z} =
0.268$, as the reference redshift. In our fiducial cosmological
framework, we have $y_s=\mes{0.147}{0.008}{0.010}$, as
  reported also in Table \ref{t|parameters}.

\begin{figure*}
  \begin{center}
    \begin{tabular}{p{0.45\textwidth} p{0.5\textwidth}}
    \vspace{0pt}
    \includegraphics[width=\linewidth]{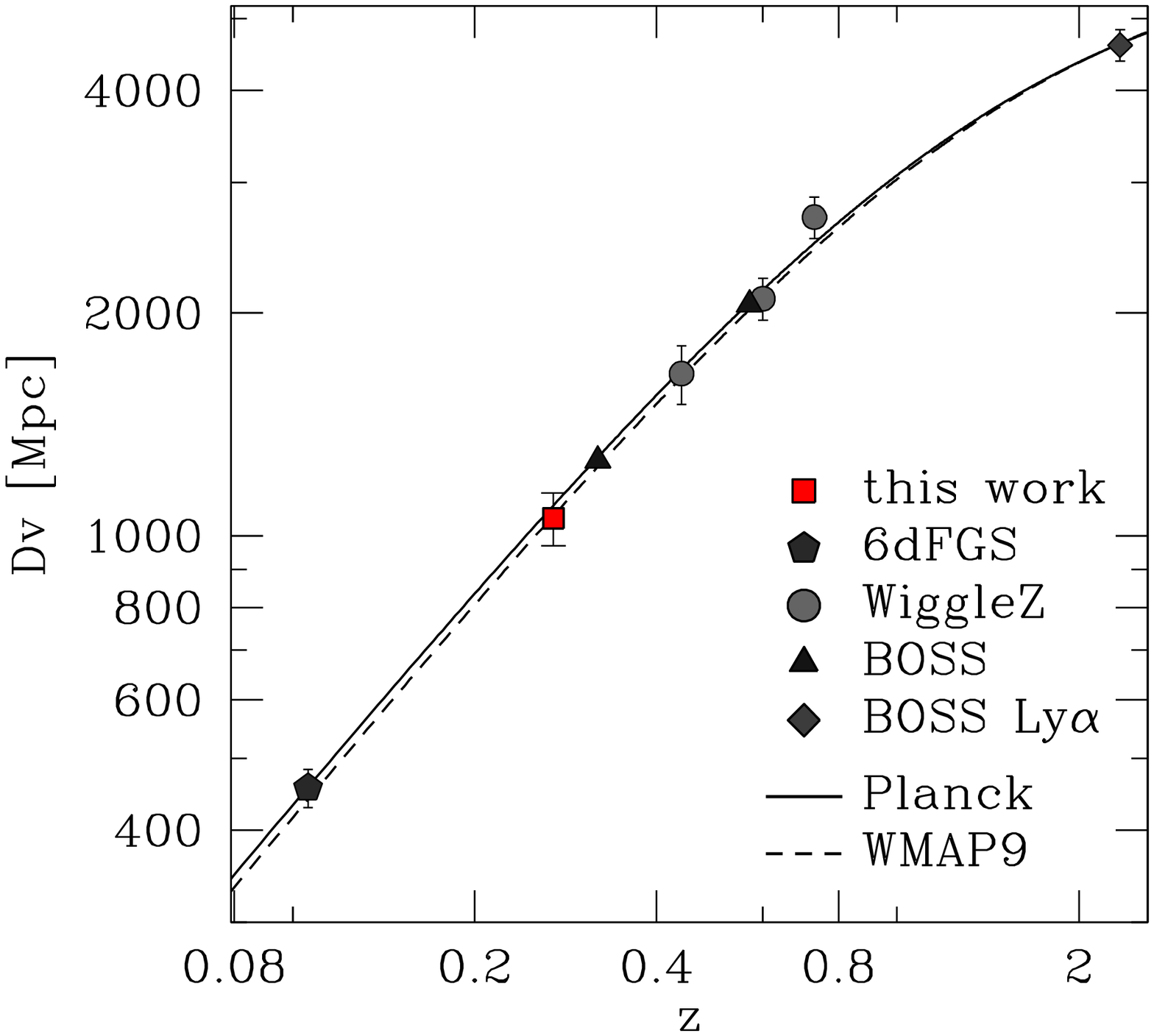} &
    \vspace{0pt}
    \includegraphics[width=0.425\textwidth]{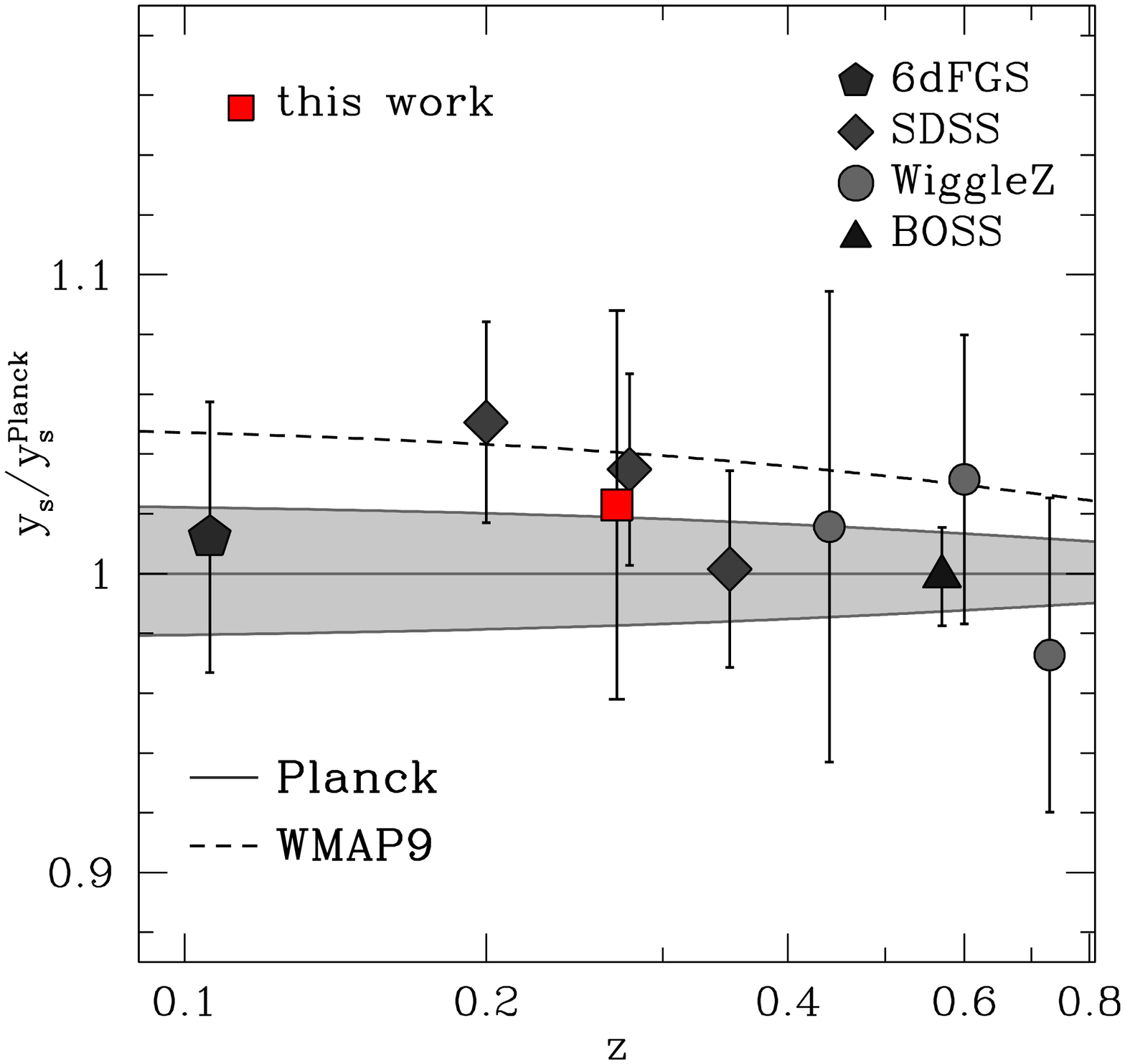}
    \end{tabular}
    \caption{$D_V$ ({\em left panel}) and $y_s$ ({\em right
        panel}) as a function of redshift. Comparison of our
      measurements (red square) with previous estimates from large
      galaxy samples (grey/black symbols): 6dF Galaxy Survey
      ($z=0.106$ by \citealt{beutler2011}), SDSS ($z=0.2, 0.35$ by
      \citealt{percival2010}; $z=0.278$ by \citealt{kazin2010}),
      WiggleZ ($z=0.44, 0.6$ and $0.73$ by \citealt{blake2011}) and
      BOSS ($z=0.35, z=0.57$ by \citealt{anderson2012}) and BOSS
      $Ly\alpha$ ($z=2.34$ by \citealt{delubac2014}).  Solid and
      dashed lines show the $\Lambda{\mathrm{CDM}}$ predictions
      obtained adopting the Planck and WMAP9 parameters, respectively.
      The $y_s(z)$ values are normalized to the Planck
        values. The shaded area is obtained changing the mass density
        parameter in the range $\{\Omega_M-0.04,\Omega_M+0.04\}$,
        where the central value is set to the Planck value
        $\Omega_M h^2 = 0.1423$.}
    \label{f|dv}
  \end{center}
\end{figure*}

In Fig.~\ref{f|dv} we compare our measurements (red squares) of
$D_V$ (left panel) and $y_s$ (right panel) with previous
estimates from large galaxy surveys at different redshifts.  The $y_s$
values are normalized to the $\Lambda\mathrm{CDM}$ prediction,
$y_s^{\rm ref}(z)$, evaluated through {\small CAMB} using the Planck
cosmological parameters \citep{planck2013}.  The shaded area
  is obtained changing the mass density parameter in the range
  ${\Omega_M-0.04,\Omega_M+0.04}$, where the central value is set to
  the Planck value $\Omega_M h^2 = 0.1423$.  All the measurements
  appear compatible with the $\Lambda{\mathrm{CDM}}$ predictions for
  both WMAP9 and Planck parameters. As it can be seen, the
  uncertainties estimated in this work are competitive with what found
  with large galaxy surveys, despite the sparseness of the
  spectroscopic cluster sample considered. However, due to the
    present uncertainties, we are not yet able to distinguish between
  the two sets of cosmological parameters given by WMAP9 and Planck.

\section{Discussion}
\label{s|disc}

The cluster centres of the spectroscopic sample analysed in
  this work are determined by the positions of the BCGs (see \S
  \ref{s|data}). Therefore, the cluster clustering presented in
  previous sections corresponds exactly to the clustering of the BCG
  sample, that is about a subsample of LRGs. The latter has obviously
  a larger level of shot-noise compared to a {\em full} LRG
  sample. Thus, it is worth wondering if there is any advantage of
  using a sparse cluster sample, instead of a larger galaxy sample,
  for BAO analyses. We address this question in \S \ref{ss|bcglrg},
  where we compare directly the two-point correlation function
  measured in a large LRG sample to the one of a subsample of
  BCGs. Finally, in \S \ref{ss|specphot} we compare the clustering of
  our spectroscopic cluster sample with a larger photometric sample,
  investigating the impact of photometric redshift errors.

\subsection{Clusters vs LRGs}
\label{ss|bcglrg}

\begin{figure*}
  \begin{center}
    \begin{tabular}{p{0.45\textwidth} p{0.5\textwidth}}
      \vspace{0pt}
      \includegraphics[width=\linewidth]{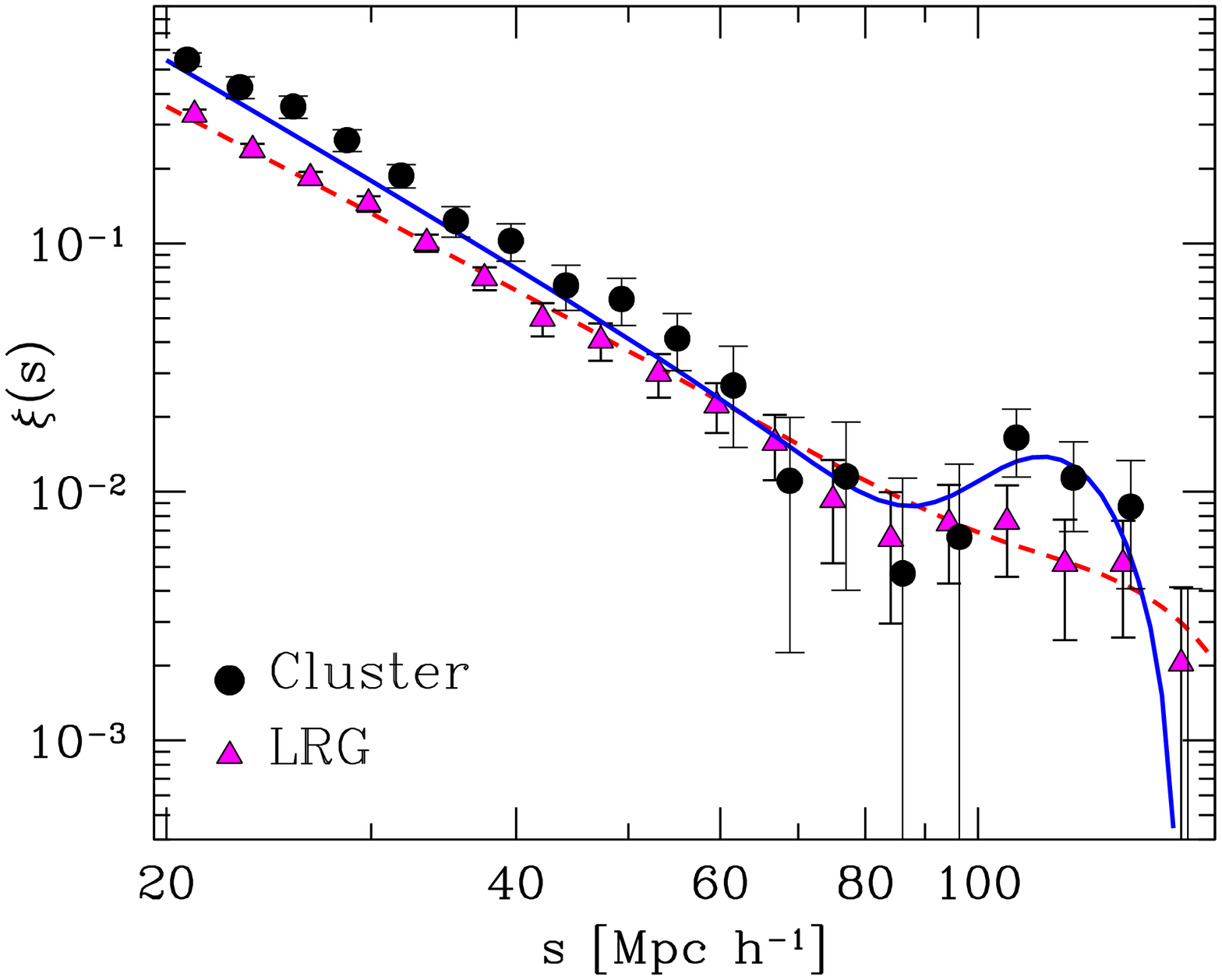} &
      \vspace{0pt}
      \includegraphics[width=0.45\textwidth]{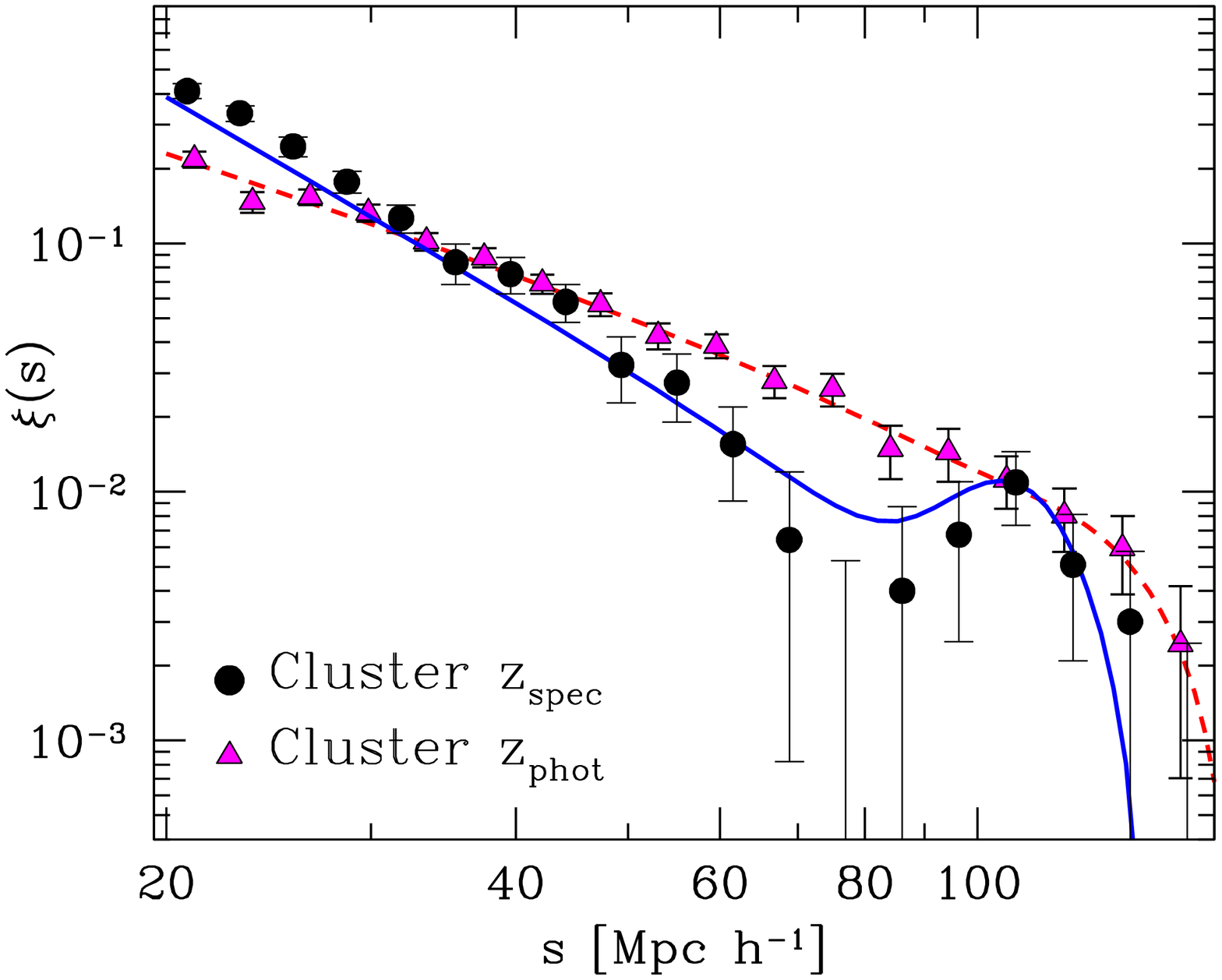}
    \end{tabular}
    \caption{Comparison between the redshift-space two-point
        correlation functions of cluster (black dots) and LRG (magenta
        triangles) spectroscopic samples ({\em left panel}), and
        between spectroscopic (black dots) and photometric (magenta
        triangles) cluster samples ({\em right panel}). The lines show
        the best-fit empirical models obtained through
        Eq.~\ref{e|sanchezmodel}, for spectroscopic clusters (blue
        lines) and photometric clusters/LRG (dashed red lines).}
    \label{f|tests}
\end{center}
\end{figure*}

For any clustering analysis, the WHL12 spectroscopic cluster
  sample can be considered just as a particularly selected subsample
  of LRGs. To investigate the impact of such a selection on the
  detection of the BAO peak, we consider here the large LRG sample by
  \citet{kazin2010}. The catalogue consists of $\sim 66000$ galaxies
  extracted from the SDSS Data Release 7, in the redshift range
  $0.16<z<0.36$. To avoid any possible systematic effect, we restrict
  our analysis to the Northern Galactic Cap, reducing the number of
  objects to $\sim 59000$, with a median redshift of
  $\tilde{z}=0.278$. Then, we identify the BCGs included in the LRG
  sample, thus obtaining the analogous of a spectroscopic cluster
  catalogue. This BCG catalogue contains $\sim 15000$ objects, with a
  median absolute magnitude larger that that of the LRG sample. The
  redshift-space two-point correlation functions of the LRG (magenta
  triangles) and BCG (black dots) samples are compared in the left
  panel of Fig.~\ref{f|tests}. The two populations show a different
  linear bias, $b_{BCG}/b_{LRG} \sim 1.16$. Moreover, while the
  $1\sigma$ error bars are smaller for the LRGs, due to their higher
  number density, the BAO peak is significantly better determined for
  the BCG sample. Following the same analysis performed in \S
  \ref{sss|baomodel}, we find that the significance of the BAO
  detection in the LRG sample is at less than $1.5\sigma$ level, and
  the error on the BAO peak results more than two times larger
  relative to the one obtained with the BCG sample. To investigate the
  robustness of our data reduction and clustering measurements, we
  compare our results with the literature \citep{kazin2010}, finding
  good agreement and confirming that our jackknife method slightly
  overpredicts the uncertainties relative to external methods based on
  mock catalogues, thus providing conservative estimates for the
  errors.
  
To further investigate the differences between BCG and LRG
  clustering, we extract two subsamples of LRGs with the same number
  of objects, one totally random and the other reproducing the BCG
  absolute magnitude distribution. Then we measure the two-point
  correlation function for both the samples, and we repeat the BAO
  analysis. In both cases, we find that the BAO peak is less
  accurately determined with respect to the BCG case. We conclude that
  BCGs, or equivalently galaxy clusters, are optimal tracers to detect the
  BAO peak. This is due to the dynamical state of these objects, that
  have significantly lower peculiar velocities with respect to other
  galaxies. Indeed, as we verified directly, the Fingers of God feature
  is almost absent in the BCG sample analysed here. This is the
  crucial property that can reduce the width of the BAO peak, thus
  improving the significance of the detection. Actually, such a small
  scale effect can directly impact the large scale clustering, as
  clearly shown in Fig.~\ref{f|tests}.


\subsection{Spectroscopic sample vs photometric sample}
\label{ss|specphot}

The effect of small scale dynamics on the two-point
  correlation function appears quite similar to the one of redshift
  errors \citep{marulli2012}. Thus, for what we have seen in \S
  \ref{ss|bcglrg}, we expect that photometric redshift errors can have
  a significant impact also at the BAO scales. To investigate this
  effect, we consider the large photometric cluster sample provided by
  WHL12, that contains more than $120000$ objects identified using a
  Friends-of-Friends algorithm. The redshift of the identified galaxy
  clusters is the mean of the photometric redshifts of their
  components. The cluster photometric redshifts result highly
  scattered around the spectroscopic redshifts, with a standard
  deviation of $\sim0.015$, as estimated in \citet{wen2012}.

The right panel of Fig.~\ref{f|tests} shows the comparison
  between the redshift-space two-point correlation function of
  photometric (magenta triangles) and spectroscopic (black dots) cluster
  samples, in the redshift range $0.1 < z < 0.42$. The number of
  photometric clusters is almost twice bigger than the spectroscopic
  one. However, as it can be seen, the large photometric redshift errors
  reduce the clustering slope \citep{marulli2012} and, most
  importantly, they broaden the BAO feature, causing a loss of information
  at the BAO scale. Indeed, the larger number of clusters does not
  compensate for the poor redshift measurements.

Finally, to further test this effect, we add Gaussian redshift
  errors to the spectroscopic sample, repeating the analysis for
  different values of the error. We find that the determination of the
  BAO peak is quite robust for redshift errors lower than $0.005$, and then
  it rapidly degrades.


\section{Conclusions}
\label{s|conclu}

In this paper we presented new measurements of the two-point
correlation function of a spectroscopic sample of galaxy clusters,
selected from the SDSS (WHL12) in the redshift range $0.1<z<0.42$.
From the projected correlation function, we derive the correlation
length and the power-law index of the real-space clustering, and the
linear bias factor. As shown in Fig.~\ref{f|xi}, we could clearly
detect the BAO peak. Fitting the measured $\xi(s)$ with an
  empirical model with a Gaussian function at the BAO scale, we find
  $s_p=\mes{104}{6}{7} \Mpch$,
  $y_s=\mes{0.147}{0.008}{0.010}$ and
  $D_V=\mes{1031}{92}{84}\Mpch$.  We test two different methods
  to analyse the BAO feature, both providing compatible
  constraints. We estimate a confidence level in the BAO detection of
  $\sim2.5\sigma$, despite the sparseness of the spectroscopic cluster sample
  considered. This is comparable to what obtained from many large galaxy surveys,
  though the latest measurements provide even stronger constraints, e.g.
  SDSS DR11 LRGs and QSO Ly-$\alpha$ provide $\sim7\sigma$ and $\sim5\sigma$ BAO
  detection, respectively \citep{anderson2014,delubac2014}.
  Overall, our measurements appear consistent with all
previous studies and with the $\Lambda\mathrm{CDM}$
predictions. Our error estimates are quite conservative,
  due to the method used to compute the covariance matrix.  The
  goodness of our results is due to the high clustering signal
  (i.e. high bias) of the cluster sample analysed, and to the accuracy
  in the spectroscopic redshift measurements. Indeed, as we have
  verified directly, the BAO peak is weakly constrained when using
  larger LRG or photometric cluster catalogues. This result shows
that galaxy clusters are powerful cosmological probes for the
detection of BAO, even with a fairly limited statistics, and highly
competitive with respect to galaxies. Future massive surveys such as
Euclid \citep[][]{laureijs2011, amendola2013} and eROSITA
  \citep{merloni2012} will allow this approach to be fully exploited
in several open key questions (e.g. the dark energy equation of
state). Accurate forecasts on the cosmological constraints
  achievable by these future cluster surveys will be provided in a
  future work. Moreover, thanks to the ongoing BOSS program, we plan
  to enrich the spectroscopic cluster sample analysed in this work,
  providing new BAO constraints at different redshifts, and as a
  function of the cluster richness.


\section*{Acknowledgments}
We would like to thank the anonymous referee for providing
  useful comments, that significantly helped to improve the paper.
We acknowledge financial contributions by grants ASI/INAF I/023/12/0,
PRIN MIUR 2010-2011 ``The dark Universe and the cosmic evolution of
baryons: from current surveys to Euclid'' and PRIN INAF 2012 ``The
Universe in the box: multiscale simulations of cosmic structure''.

\label{lastpage}

\end{document}